\overfullrule=0pt
\input harvmac
\def\a{{\alpha}}
\def\ah{{\widehat\alpha}}

\def\l{{\lambda}}
\def\lh{{\widehat\lambda}}
\def\lt{{\widetilde\lambda}}
\def\b{{\beta}}
\def\bh{{\widehat\beta}}
\def\g{{\gamma}}
\def\gh{{\widehat\gamma}}
\def\d{{\delta}}
\def\dh{{\widehat\delta}}

\def\s{{\sigma}}

\def\s{{\sigma}}

\def\N{{\nabla}}
\def\Nb{{\overline\nabla}}
\def\Nh{{\widehat\nabla}}
\def\O{{\Omega}}

\def\Oh{{\widehat\O}}
\def\o{{\omega}}
\def\oh{{\widehat\omega}}
\def\ot{{\widetilde\omega}}
\def\half{{1\over 2}}
\def\p{{\partial}}
\def\pb{{\overline\partial}}
\def\t{{\theta}}

\def\oh{{\widehat\o}}

\def\dhh{{\widehat d}}

\def\Pib{{\overline\Pi}}
\def\Jb{{\overline J}}

\def\S{{\Sigma}}
\def\Sh{{\widehat\Sigma}}
\def\dhh{{\widehat d}}
\def\Ch{\widehat C}
\def\Rh{\widehat R} 
\def\lth{{\widehat\lt}}
\def\oth{{\widehat\ot}}
\def\rhh{{\widehat{r}}}
\def\shh{{\widehat{s}}}
\def\Ah{{\widehat{A}}}
\def\Str{{\rm Str}}

\baselineskip12pt

\Title{ \vbox{\baselineskip12pt
}}
{\vbox{\centerline
{Non-minimal fields of the pure spinor string }
\bigskip
\centerline{in general curved backgrounds }
}}
\smallskip
\centerline{Osvaldo Chandia\foot{e-mail: ochandiaq@gmail.com}, }
\smallskip
\centerline{\it Departamento de Ciencias, Facultad de Artes Liberales, Universidad Adolfo Ib\'a\~nez}
\centerline{\it Facultad de Ingenier\'{\i}a y Ciencias, Universidad Adolfo Ib\'a\~nez}
\centerline{\it Diagonal Las Torres 2640, Pe\~nalol\'en, Santiago, Chile} 
\bigskip
\centerline{Brenno Carlini Vallilo\foot{e-mail: vallilo@gmail.com}, }
\smallskip
\centerline{\it Departamento de Ciencias F\'{\i}sicas, Facultad de Ciencias Exactas}
\centerline{\it Universidad Andr\'es Bello, Rep\'ublica 220, Santiago, Chile}

\vskip 0.8in

\noindent
We study the coupling of the non-minimal ghost fields of the pure spinor superstring 
in general curved backgrounds. The coupling is found solving the consistent relations 
from the nilpotency of the non-minimal BRST charge.

\Date{December 2014}


\lref\BerkovitsFE{
  N.~Berkovits,
  ``Super Poincare covariant quantization of the superstring,''
JHEP {\bf 0004}, 018 (2000).
[hep-th/0001035].
}

\lref\BerkovitsNG{
  N.~Berkovits and C.~R.~Mafra,
  ``Equivalence of two-loop superstring amplitudes in the pure spinor and RNS formalisms,''
Phys.\ Rev.\ Lett.\  {\bf 96}, 011602 (2006).
[hep-th/0509234].
}

\lref\BerkovitsIC{
  N.~Berkovits and J.~Maldacena,
  ``Fermionic T-Duality, Dual Superconformal Symmetry, and the Amplitude/Wilson Loop Connection,''
JHEP {\bf 0809}, 062 (2008).
[arXiv:0807.3196 [hep-th]].
}

\lref\ChandiaYV{
  O.~Chandia,
  ``A Note on T-dualities in the Pure Spinor Heterotic String,''
JHEP {\bf 0904}, 104 (2009).
[arXiv:0902.2729 [hep-th]].
}

\lref\BerkovitsXU{
  N.~Berkovits,
  ``Quantum consistency of the superstring in AdS(5) x S**5 background,''
JHEP {\bf 0503}, 041 (2005).
[hep-th/0411170].
}

\lref\ValliloFJ{
  B.~C.~Vallilo and L.~Mazzucato,
  ``The Konishi multiplet at strong coupling,''
JHEP {\bf 1112}, 029 (2011).
[arXiv:1102.1219 [hep-th]].
}

\lref\ChandiaWCA{
  O.~Chandia, L.~I.~Bevilaqua and B.~C.~Vallilo,
  ``AdS pure spinor superstring in constant backgrounds,''
JHEP {\bf 1406}, 029 (2014).
[arXiv:1404.0974 [hep-th]].
}

\lref\BerkovitsAIA{
  N.~Berkovits,
  ``Twistor Origin of the Superstring,''
[arXiv:1409.2510 [hep-th]].
}

\lref\BerkovitsPX{
  N.~Berkovits,
  ``Multiloop amplitudes and vanishing theorems using the pure spinor formalism for the superstring,''
JHEP {\bf 0409}, 047 (2004).
[hep-th/0406055].
}

\lref\BerkovitsBT{
  N.~Berkovits,
  ``Pure spinor formalism as an N=2 topological string,''
JHEP {\bf 0510}, 089 (2005).
[hep-th/0509120].
}

\lref\BerkovitsVI{
  N.~Berkovits and N.~Nekrasov,
  ``Multiloop superstring amplitudes from non-minimal pure spinor formalism,''
JHEP {\bf 0612}, 029 (2006).
[hep-th/0609012].
}

\lref\BakhmatovFPA{
  I.~Bakhmatov and N.~Berkovits,
  ``Pure Spinor $b$-ghost in a Super-Maxwell Background,''
JHEP {\bf 1311}, 214 (2013).
[arXiv:1310.3379 [hep-th]].
}

\lref\ChandiaIMA{
  O.~Chandia,
  ``The Non-minimal Heterotic Pure Spinor String in a Curved Background,''
[arXiv:1311.7012 [hep-th]].
}

\lref\BerkovitsUE{
  N.~Berkovits and P.~S.~Howe,
  ``Ten-dimensional supergravity constraints from the pure spinor formalism for the superstring,''
Nucl.\ Phys.\ B {\bf 635}, 75 (2002).
[hep-th/0112160].
}

\lref\ChandiaHN{
  O.~Chandia and B.~C.~Vallilo,
  ``Conformal invariance of the pure spinor superstring in a curved background,''
JHEP {\bf 0404}, 041 (2004).
[hep-th/0401226].
}

\lref\ChandiaIX{
  O.~Chandia,
  ``A Note on the classical BRST symmetry of the pure spinor string in a curved background,''
JHEP {\bf 0607}, 019 (2006).
[hep-th/0604115].
}

\lref\BedoyaIC{
  O.~A.~Bedoya and O.~Chandia,
  ``One-loop Conformal Invariance of the Type II Pure Spinor Superstring in a Curved Background,''
JHEP {\bf 0701}, 042 (2007).
[hep-th/0609161].
}

\lref\BerkovitsYR{
  N.~Berkovits and O.~Chandia,
  ``Superstring vertex operators in an AdS(5) x S**5 background,''
Nucl.\ Phys.\ B {\bf 596}, 185 (2001).
[hep-th/0009168].
}

\lref\MetsaevIT{
  R.~R.~Metsaev and A.~A.~Tseytlin,
  ``Type IIB superstring action in AdS(5) x S**5 background,''
Nucl.\ Phys.\ B {\bf 533}, 109 (1998).
[hep-th/9805028].
}

\lref\BerkovitsPLA{
  N.~Berkovits,
  ``Dynamical twisting and the b ghost in the pure spinor formalism,''
JHEP {\bf 1306}, 091 (2013).
[arXiv:1305.0693 [hep-th]].
}

\lref\BerkovitsAMA{
  N.~Berkovits and O.~Chandia,
  ``Simplified Pure Spinor $b$ Ghost in a Curved Heterotic Superstring Background,''
JHEP {\bf 1406}, 001 (2014).
[arXiv:1403.2429 [hep-th]].
}

\lref\MikhailovRX{
  A.~Mikhailov,
  ``Symmetries of massless vertex operators in AdS(5) x S**5,''
J.\ Geom.\ Phys.\  {\bf 62}, 479 (2012).
[arXiv:0903.5022 [hep-th]].
}


\newsec{ÊIntroduction}

The pure spinor formalism was introduced by Berkovits in \BerkovitsFE\ as an alternative 
to RNS and Green-Schwarz formalisms. Its main advantages are manifest space time supersymmetry and the fact that 
it can be quantized in the conformal gauge. Since its introduction, the formalism was used in wide range of applications, 
including the computation of loop scattering amplitudes \BerkovitsNG, fermionic T-Duality \refs{\BerkovitsIC,  \ChandiaYV} and quantum computations in $AdS$ background \refs{\BerkovitsXU,\ValliloFJ,\ChandiaWCA}. All the results obtained with the formalism are compatible with RNS or GS when the comparison is possible. 

However, many aspects of the formalism remain elusive. For example, it is not yet known which action one applies  
the standard BRST gauge fixing to obtain pure spinor action \foot{For a different approach see \BerkovitsAIA.}. 
A related problem is that we do not have a pure spinor formalism out of the conformal gauge. As a consequence, 
the usual conformal ghosts $(b,c)$ are not fundamental fields but are composite fields constructed out spacetime and 
ghost fields. Although it is possible to define multiloop amplitudes with only the minimal variables \BerkovitsPX, the non-minimal variables \BerkovitsBT\ were introduced to have a better understanding of the construction of the $b$ ghost, which can be seen as one of the generators of a twisted $N=2$ superconformal algebra. Later, the non-minimal formulation was used to define multiloop amplitudes in \BerkovitsVI.

A natural question to ask is how to generalize the non-minimal pure spinor string to curved backgrounds. This was first investigated by Bakhmatov and Berkovits \BakhmatovFPA, where the $b$ ghost was constructed for a super-Maxwell background in the open superstring. Later, one of the authors of the present paper derived \ChandiaIMA\ the coupling of the non-minimal fields in a curved heterotic background and also initiated the construction of the $b$ ghost in this background. In this paper we construct the pure spinor BRST transformations of the non-minimal fields in a general curved background by demanding nilpotency of the BRST charge. Although all these coupling are BRST trivial by construction, it is much better to have a formulation of the theory where all local symmetries are manifest, so the non-minimal fields couple to the spin connection. 

This paper is organized as follows. In section 2, we summarize the work done in \ChandiaIMA\ and discuss the modification of the approach used in this reference that will be used in the following sections. In section 3 we apply the new approach for the type II string. The especial case of $AdS_5 \times S^5$ background is done Section 4. This section was intentionally written in a way that can be read independently of the previous sections. In the last section we discuss possible further developments. 
 
\newsec{ Non-minimal BRST Transformations}

The minimal pure spinor string on a flat background is defined by a free-field action for the ten-dimensional superspace coordinates $(X^m , \t^\a, p_\a)$, where $p_\a$ is the momentum conjugate of $\t^\a$. As it was noted by Berkovits \BerkovitsFE, the resulting action is not conformal invariant and introduced a pure spinor variable $\l^\a$ which is constrained by $(\l \g^m \l) = 0$, where $(\g^m)_{\a\b}$ are the $16 \times 16$ symmetric Pauli matrices in ten dimensions. The BRST charge is $Q = \oint \l^\a d_\a$, where $d_\a$ is the world-sheet generator for the superspace covariant derivative. Note that $Q$ is nilpotent because the OPE algebra of $d_\a$ and the pure spinor condition. 

The BRST charge reproduces the superstring spectrum and it helps to define correctly the superstring tree-level scattering amplitudes. To define loop scattering amplitudes it was necessary to include non-minimal pure spinor variables. In order to include new variables  without changing the spectrum, their BRST charge has to have trivial cohomology. This is done by adding the bosonic conjugate pair $(\lt_\a, \ot^\a)$ and the fermonic conjugate pair $(r_\a , s^\a)$. Their contribution to the BRST charge is $\oint \ot^\a r_\a$ which has trivial cohomology \BerkovitsBT.  Up to now, we have discussed the left-moving sector of the pure spinor string. If we are interested in the heterotic string, we add the right-moving sector corresponding to the bosonic string. Similarly, if we are interested in the type II string, we add the right-moving sector with the same chirality of the left-moving sector for the type IIB string, or  with the opposite chirality for the type IIA string.

We now discuss the case of the heterotic string in a curved background. Consider the left-moving sector. The superspace coordinates are $Z^M$ and the pure spinor variable is coupled to the background by changing $\pb \l^\a$ to 
\eqn\couplingpure{\Nb \l^\a = \pb \l^\a + \l^\b \pb Z^M \O_{M \b}{}^\a ,} 
where $\O_{M\b}{}^\a$ is the background connection and it has to be of the form
\eqn\conn{ \O_{M\b}{}^\a = \O_M \d_\b^\a + {1\over 4} \O_{M ab} (\g^{ab})_\b{}^\a ,}
to preserve the gauge invariance $\delta \omega_\a= (\gamma^a\lambda)_\a\Lambda_a$. This gauge invariance is a consequence of the pure spinor condition. Note that $\O_\a = E_\a{}^M \O_M$ ($E$ is the vielbein super field) is related to the super-dilaton field, $\Phi$,  which defines the Fradkin-Tseytlin term in the action. In fact, such relation comes from ghost number anomaly cancellation \BerkovitsUE.  The precise relations is $4\N_\a \Phi = \O_\a$. Note that this relation is also obtained from the computation of the vanishing of the one-loop beta functions \ChandiaHN. The superspace metric is given in terms of the vielbein superfield  $E_M{}^A$ and its inverse $E_A{}^M$ as    
\eqn\metric{ G_{NM}(Z) = E_N{}^a E_M{}^b \eta_{ab} ,}
where $\eta$ is the Minkowski metric in ten dimensions.   

The superstring action has the form
\eqn\smc{ S = \int d^2z ~ \left( \ha \p Z^M \pb Z^N G_{NM}(Z) + \o_\a \Nb \l^\a + d_\a \pb Z^M E_M{}^\a + \cdots \right) ,}Êwhere the terms $\cdots$ are determined by demanding that the expansion of \smc\ around flat background gives the massless integrated vertex operator \BerkovitsUE. Equivalently, one could write the most general action which is classically conformal invariant and then define a BRST charge

\eqn\Qbrst{Q = \oint \l^\a d_\a ,}
which is nilpotent and conserved. This process requires the background satisfy some constraints. It turns out that these constraints are solved by 
\eqn\torsioncomps{ T_{a\a}{}^\b = T_{\a\b}{}^\g = 0,\quad T_{\a\b}{}^a = \g^a_{\a\b} ,}
where $T^A = (T^a, T^\a)$ is the torsion two-form defined as the exterior covariant derivative of the vielbein one-form $E^A = dZ^M E_M{}^A$, that is
\eqn\torsion{ T^A = \N E^A = d E^A + E^B \O_B{}^A ,} 
 where $\O_B{}^A = dZ^M \O_{MB}{}^A$ is the connection one-form. The ten-dimensional supergravity equations of motion are obtained from \torsioncomps\ and the Bianchi identities
\eqn\bianchiT{ \N T^A = T^B R_B{}^A ,} 
where $R_B{}^A$ is the curvature 2-form given by $R_B{}^A = d \O_B{}^A + \O_B{}^C \O_C{}^A$.

A natural question is how does the world-sheet fields of the action \smc\ change under the action of the BRST charge \Qbrst\ and verify that the action is invariant. This was done in \ChandiaIX\ where the transformations were obtained by expressing the world-sheet field $d_\a$ in terms of the remaining world-sheet fields of \smc\ and the momentum conjugate of $Z^M$, and using canonical commutation relations. The resulting BRST transformations are
\eqn\Qzero{ Q \l^\a = -\l^\b \left( \l^\g \O_{\g\b}{}^\a \right),\quad Q \o_\a = d_\a +  \left( \l^\g \O_{\g\a}{}^\b \right) \o_\b ,}
$$ Q d_\a = \Pi^a (\g_a\l)_\a + \l^\b R_{\b\a\g}{}^\d \l^\g \o_\d + \left( \l^\g \O_{\g\a}{}^\b \right) d_\b ,$$
where $\Pi^a = \p Z^M E_M{}^a$. We can define $\Pi^\a = \p Z^M E_M{}^\a$ and determine the BRST transformation of $\Pi^A$ from $Q Z^M = \l^\a E_\a{}^M$. The result is
\eqn\Qpis{ÊQ \Pi^a = - \l^\a \Pi^A T_{A\a}{}^a - \Pi^b \left( \l^\g \O_{\g b}{}^a \right),\quad Q \Pi^\a = \N \l^\a + \Pi^\b  \left( \l^\g \O_{\g\b}{}^\a \right) .}
Note that in the BRST transformations of \Qzero\ and \Qpis\ appears a term which is a Lorentz rotation with the field-dependent parameter $\S_\b{}^\a = \l^\g \O_{\g\b}{}^\a$. Every field that carries a Lorentz index should contain a Lorentz rotation of this type in its BRST transformation. In particular, the non-minimal fields are not inert under Lorentz rotations. Then, it is expected that \Qbrst\ acts non-trivially on them. It was determined in \ChandiaIMA\ the form in which the non-minimal fields are transformed under the minimal BRST charge. The idea is to impose that \Qbrst\ is nilpotent on them and that the minimal sector is not affected by the non-minimal BRST charge $\oint \ot^\a r_\a$.  After adding the non-minimal BRST charge to \Qbrst, it was obtained that
\eqn\qnmhet{ Q r_\a = -{1\over4} \l^\g T_{\g ab} (\g^{ab})_\a{}^\b r_\b + \S_\a{}^\b r_\b,\quad Q s^\a = \ot^\a - {1\over4} s^\b \l^\g T_{\g ab} (\g^{ab})_\b{}^\a + s^\b \S_\b{}^\a,}
$$ Q \lt_\a = -r_\a - {1\over4} \l^\g T_{\g ab} (\g^{ab})_\a{}^\b \lt_\b + \S_\a{}^\b \lt_\b,\quad Q \ot^\a =  - {1\over4} \ot^\b \l^\g T_{\g ab} (\g^{ab})_\b{}^\a - \ot^\b \S_\b{}^\a .$$
This solution is only viable when the torsion $T_{\a ab}$ does not vanish which is required by the nilpotency of $Q_0$ on the non-minimal fields. One important background with vanishing $T_{\a ab}$ is $AdS_5\times S^5$, so it is relevant to determine the BRST transformations of the non-minimal fields in this case and in a generic type II background. It will be shown that it is not possible to keep nilpotency of $Q_0$ and vanishing $T_{\a ab}$ simultaneously. Therefore, we apply a different strategy. 

We now construct a different set of BRST transformations for the non-minimal fields which will be useful for the type II superstring. Instead of imposing nilpotency of the minimal BRST on the non-minimal variables, we impose nilpotency of the full BRST charge. Note that the BRST transformations still have to contain the field-dependent Lorentz rotation piece.  We fix the form of other possible contributions to the BRST transformations by counting ghost number and then computing $Q^2$ on the non-minimal fields. Consider $\lt$ and $r$. The non-minimal BRST charge relates the ghost number of $\lt$, which we call $q_\lt$, to the ghost number of $r$, which we call $q_r$. Since the BRST charge carries ghost number one and the fact that the non-minimal BRST charge maps $\lt$ into $-r$, we obtain that $q_r - q_\lt =1$.  They have conformal weight zero, then any term linear in $\ot$ and $s$ should be multiplied by terms that depend on the minimal variables and have conformal weight $-1$. We discard this option because is not possible to construct such fields. For $\lt$, the action of the non-minimal BRST charge already determines a term with $r$, the other possible term is the Lorentz rotation. Then we have
\eqn\qlt{ Q \lt_\a = - r_\a + \S_\a{}^\b \lt_\b  .} 
Consider now $r$. The non-minimal BRST charge annihilates it and a Lorentz rotation has to be present. A new possible contribution should be linear in $\lt$. Because $Q r$ has ghost number $1+q_r = 2 + q_\lt$, the term multiplying $\lt$ has ghost number $2$. That is,
\eqn\qr{ÊQ r_\a = A_\a{}^\b \lt_\b + \S_\a{}^\b r_\b ,}
where $A$ is an even field with ghost number $2$ and depends on the minimal fields only. Similarly, the other non-minimal fields transform like
\eqn\qsqot{ÊQ s^\a = \ot^\a + s^\b \S_\b{}^\a,\quad Q \ot^\a = s^\b A_\b{}^\a - \ot^\b \S_\b{}^\a .}
We determine $A$ through the equation $Q^2=0$. When $Q$ acts twicelly on $\lt$ we obtain
\eqn\detA{ A_\a{}^\b = Q \S_\a{}^\b - \S_\a{}^\g \S_\g{}^\b .}Ê
The same equation is obtained when we compute $Q^2 s^\a = 0$. The equation $Q^2 r_\a = 0$ we obtain
\eqn\QA{ÊQ A_\a{}^\b + A_\a{}^\g \S_\g{}^\b - \S_\a{}^\g A_\g{}^\b = 0 ,}
which is satisfied by acting with $Q$ in the equation \detA. The  same equation is obtained from $Q^2 \ot^\a=0$. Remembering that $\S_\a{}^\b = \l^\g \O_{\g\a}{}^\b$ and the minimal BRST transformations, $A$ from \detA\ turns out to be proportional to the curvature, in fact
\eqn\AisR{ÊA_\a{}^\b = \ha \l^\g \l^\d R_{\g\d\a}{}^\b .}
In summary, the BRST transformation of the non-minimal variables are
\eqn\qnmhetp{ Q r_\a = \ha \l^\g \l^\d R_{\g\d\a}{}^\b \lt_\b + \S_\a{}^\b r_\b,\quad Q s^\a = \ot^\a + s^\b \S_\b{}^\a ,}
$$ Q \lt_\a = - r_\a + \S_\a{}^\b \lt_\b,\quad Q \ot^\a = \ha s^\b \l^\g \l^\d R_{\g\d\b}{}^\a  - \ot^\b \S_\b{}^\a .$$ 

Having in mind the case of the superstring on $AdS_5 \times S^5$ background, the expressions \qnmhetp\ seems more appropriate than \qnmhet\ because on this background the dilation is constant and, consequently, the torsion $T_{\a ab}$ vanishes and the transformations in \qnmhet\ are not nilpotent. We now generalize the heterotic string result of \qnmhetp\ to the type II superstring and then to the type IIB superstring on $AdS_5 \times S^5$ background.    

\newsec{Type II superstring in the minimal and non-minimal formulation}

We generalize the above result for the type II superstring. We first discuss the minimal sector and then we add the non-minimal variables. In the minimal sector, we determine the the BRST transformation of the fields following \ChandiaIX. 

\subsec{ The minimal type II superstring}

The action in the minimal type II superstring is
\eqn\Stypeii{ S = \int d^2 z ~ \ha \Pi^a \Pib^b \eta_{ab} + \ha \Pi^A \Pib^B  B_{BA} + d_\a \Pib^\a + \dhh_\ah \Pi^\ah  + d_\a \dhh_\bh P^{\a\bh} }
$$ + \l^\a \o_\b 	\dhh_\gh C_\a{}^{\b\gh} + \lh^\ah \oh_\bh d_\g \Ch_\ah{}^{\bh\g} + \l^\a \o_\b \lh^\gh \oh_\dh S_{\a\gh}{}^{\b\dh} + \o_\a \Nb \l^\a + \oh_\ah \N \lh^\ah +S_{FT} ,$$
where besides the vielbein and the spin connection, the background fields are the Kalb-Ramond superfield $B$,  the superfield $P$ is the Ramond-Ramond field strength, the superfields $C$ and $\Ch$ contain the graviton and the dilatini,  and $S$ is related to the curvature of the background super geometry. $S_{FT}$ is the Fradkin-Tseytlin superfield which depends on the dilation superfield $\Phi$. This term is important to preserve quantum conformal invariance  \BedoyaIC. 
 Note that $\Pi^A = \p Z^M E_M{}^A = ( \Pi^a , \Pi^\a , \Pi^\ah )$ so the capital index $A$ now runs over ten-dimensional type II superspace. The background fields in \Stypeii\ are constrained by the BRST charge
 \eqn\qtypeii{ÊQ = \oint (\l^\a d_\a + \lh^\ah \dhh_\ah ).} 
It was shown in \BerkovitsUE\ that the nilpotency and the conservation of the BRST charge constrain the background fields of the action \Stypeii. Some of these constraints are solved by
\eqn\val{ T_{\a\b}{}^\g = T_{\a\bh}{}^\g = T_{\ah\bh}{}^\g = 0, \quad T_{\a\b}{}^\gh = T_{\a\bh}{}^\gh = T_{\ah\bh}{}^\gh = 0,}
$$ H_{\a\b\g} = H_{\a\b\gh} = H_{\a\bh\gh} = H_{\ah\bh\gh} = 0,\quad T_{\a\bh}{}^a = 0,\quad R_{\ah\bh\g}{}^\d = \Rh_{\a\b\gh}{}^\dh = 0 ,$$
$$ T_{\a\b a} = H_{\a\b a} = (\g_a)_{\a\b},\quad T_{\ah\bh a} = - H_{\ah\bh a} = (\g_a)_{\ah\bh} ,$$
where $H=dB$, $T^\a$ is the torsion defined with $\O$ as connection, $T^\ah$ is the torsion defined with $\Oh$ as connection. Similarly, $R$ is the curvature for the connection $\O$ and $\Rh$ is the curvature for the connection $\Oh$. The other constraints derived from the nilpotency of the BRST charge and the conservation of the BRST current allow, together with the Bianchi identities, to show that the background satisfy the type II supergravity equations in ten dimensions \BerkovitsUE. 

We now compute the BRST transformations of the world-sheet fields in \Stypeii. We follow the procedure of \ChandiaIX\  where the BRST transformations of the world-sheet heterotic fields were determined. The idea is to define the conjugate variables and then use canonical commutation relations for the world-sheet fields. Note that we already have two pairs of canonical variables in \Stypeii. They are $(\o, \l)$ and $(\oh, \lh)$ and their equal world-sheet time commutation relations are defined to be \foot{These commutation relations are not compatible with the pure spinor condition. However, if $\omega_\a$ and $\oh_\ah$ only appear in gauge invariant combinations such as $\l^\a\o_\a$ and $N^{ab}$ then we can use these commutators.}
\eqn\comma{ [ \l^\a (\s) , \o_\b(\s') ] = \d^\a_\b \d(\s-\s'),\quad [ \lh^\ah(\s) , \oh_\bh(\s') ] = \d_\bh^\ah \d(\s-\s') .} 
We define the canonical conjugate variable of $Z^M$ as
\eqn\defP{ÊP_M = {{\d S_0}\over{\d \p_0 Z^M}} ,}
where we have defined the world-sheet coordinates such that $\p = \p_0 - \p_1$ and $\pb = \p_0 + \p_1$. In this way, we obtain
\eqn\PM{ P_M = - E_M{}^\a d_\a - E_M{}^\ah \dhh_\ah + \p_0  Z^N G_{NM} + \p_1 Z^N B_{NM} + \O_{M\a}{}^\b \l^\a \o_\b + \Oh_{M\ah}{}^\bh \lh^\ah \oh_\bh ,}
where $G_{NM} = E_N{}^a E_M{}^b \eta_{ab}$ is the super metric and $B_{NM}$ is the target superspace Kalb-Ramond two form. The equal world-sheet time canonical commutator between $Z^M$ and its momentum is
\eqn\commb{Ê[ P_M(\s) ,Z^N(\s') ] = -\d_M^N \d(\s-\s') .}

We express the BRST charge \qtypeii\ in terms of the canonical variables of the action \Stypeii\ and then we compute the BRST transformations of the world-sheet fields using \comma\ and \commb. Note that $d_\a$ and $\dhh_\ah$ can be obtained from \PM\ after multiplying by the appropriate inverse of the super vielbein. That is,
\eqn\des{Êd_\a = -E_\a{}^M P_M + \p_1 Z^M B_{M\a} + \O_{\a\b}{}^\g \l^\b \o_\g + \Oh_{\a\bh}{}^\gh \lh^\bh \oh_\gh ,}  
$$ \dhh_\ah = -E_\ah{}^M P_M + \p_1 Z^M B_{M\ah} + \O_{\ah\b}{}^\g \l^\b \o_\g + \Oh_{\ah\bh}{}^\gh \lh^\bh \oh_\gh .$$

After using \des\ in \qtypeii, the BRST transformations of the fields are
\eqn\Qmin{ÊQ d_\a = ( \l^\g \O_{\g\a}{}^\b + \lh^\gh \O_{\gh\a}{}^\b ) d_\b + ( \l^\b R_{\b\a\g}{}^\d + \lh^\bh R_{\bh\a\g}{}^\d ) \l^\g \o_\d + (\g_a \l)_\a \Pi^a ,} 
$$ Q \dhh_\ah = ( \l^\g \Oh_{\g\ah}{}^\bh + \lh^\gh \Oh_{\gh\ah}{}^\bh ) \dhh_\bh + ( \l^\b \Rh_{\b\ah\gh}{}^\dh + \lh^\bh \Rh_{\bh\ah\gh}{}^\dh ) \lh^\gh \oh_\dh - (\g_a \lh)_\ah \Pib^a ,$$
$$ Q \l^\a = - \l^\b ( \l^\g \O_{\g\b}{}^\a + \lh^\gh \O_{\gh\b}{}^\a ),\quad Q \o_\a = d_\a + (\l^\g \O_{\g\a}{}^\b + \lh^\gh \O_{\gh\a}{}^\b) \o_\b ,$$
$$ Q \lh^\ah = - \lh^\bh ( \l^\g \Oh_{\g\bh}{}^\ah + \lh^\gh \Oh_{\gh\bh}{}^\ah ),\quad Q \oh_\ah = \dhh_\ah + (\l^\g \Oh_{\g\ah}{}^\bh + \lh^\gh \Oh_{\gh\ah}{}^\bh) \oh_\bh .$$
The superspace coordinate $Z^M$ transforms as $Q Z^M = \l^\a E_\a{}^M + \lh^\ah E_\ah{}^M$. Then, $\Pi^A = \p Z^M E_M{}^A$ transforms as
\eqn\qpis{ Q \Pi^a = - \l^\a \Pi^A T_{A\a}{}^a - \lh^\ah \Pi^A T_{A\ah}{}^a - \Pi^b ( \l^\a \O_{\a b}{}^a + \lh^\ah \O_{\ah b}{}^a) ,}
$$ Q \Pi^\a = \N \l^\a - \lh^\bh \Pi^a T_{a\bh}{}^\a + \Pi^\b (\l^\g \O_{\g\b}{}^\a + \lh^\gh \O_{\gh\b}{}^\a ) ,$$
$$  Q \Pi^\ah = \N \lh^\ah - \l^\b \Pi^a T_{a\b}{}^\ah + \Pi^\bh (\l^\g \Oh_{\g\bh}{}^\ah + \lh^\gh \Oh_{\gh\bh}{}^\ah ) .$$
Similarly for $\Pib^A = \pb Z^M E_M{}^A$. As in the heterotic string case \ChandiaIX, the above BRST transformations contain a Lorentz transformation with a parameter that depends on the pure spinor ghosts $\l$ and $\lh$ which we call
\eqn\sigmas{Ê\S_\a{}^\b = \l^\g \O_{\g\a}{}^\b + \lh^\gh \O_{\gh\a}{}^\b,\quad \Sh_\ah{}^\bh = \l^\g \Oh_{\g\ah}{}^\bh + \lh^\gh \Oh_{\gh\ah}{}^\bh .} 
We now construct the BRST transformations of the non-minimal pure spinor fields of the form \qnmhetp.

\subsec{ The non-minimal type II superstring}

The non minimal pure spinor variables are the left-moving ghosts $(\lt_\a, \ot^\a, r_\a, s^\a)$ and the right-moving ghosts $(\lth_\ah, \oth^\ah, \rhh_\ah, \shh^\a)$. Their contribution to the BRST charge is 
\eqn\qnmtypeii{ÊQ_1 = \oint (\ot^\a r_\a + \oth^\a \rhh_\a) .} 

First, let us consider the left-moving sector. The BRST charge \qnmtypeii\ acts on them as
\eqn\qone{ÊQ_1 \lt_\a = - r_\a,\quad Q_1 s^\a = \ot^\a,\quad Q_1 \ot^\a = Q_1 r_\a = 0 .} 
As it was noted in \ChandiaIMA, the non-minimal pure spinor variables are not inert under the action of the minimal pure spinor charge of \qtypeii.  We now find the action of \qtypeii\ on the non-minimal pure spinor fields by imposing that the full BRST charge, the sum of \qtypeii\ and \qnmtypeii, is nilpotent. As in the heterotic string case, the equations \qone\ are assumed to be true in any type II curved background. Consider first $r$ and $\lt$. Following the ghost number analysis of above, the action of \qtypeii\ on them is such that the BRST transformations of these non-minimal fields have the form
\eqn\Qrl{ Q r_\a = \S_\a{}^\b r_\b + A_\a{}^\b \lt_\b,\quad Q \lt_\a = - r_\a + \S_\a{}^\b \lt_\b ,}
where $A$ is an undetermined function of the minimal variables with ghost number $2$. Imposing $Q^2 \lt_\a = 0$, implies that
\eqn\Ais{ÊA_\a{}^\b = Q \S_\a{}^\b - \S_\a{}^\g \S_\g{}^\b .}
We now impose $Q^2 r_\a = 0$ which implies that $A$ is given by \Ais\ and the equation
\eqn\QA{ÊQ A_\a{}^\b + A_\a{}^\g \S_\g{}^\b - \S_\a{}^\g A_\g{}^\b = 0 .}
Note that this equation is obtained from \Ais\ after applying $Q$.  Consider now $s$ and $\ot$. It turns out that $Q$ on them is
\eqn\Qso{ Q s^\a = \ot^\a + s^\b \S_\b{}^\a,\quad Q \ot^\a = - \ot^\b \S_\b{}^\a + s^\b A_\b{}^\a ,} 
with the same function $A$ \Ais. Computing the r.h.s. of \Ais\ implies that
\eqn\Ares{ÊA_\a{}^\b = \ha \l^\g \l^\d R_{\g\d\a}{}^\b + \l^\g \lh^\dh R_{\g\dh\a}{}^\b  .}

We proceed similarly for the non-minimal right-moving sector $(\lth , \oth), (\rhh , \shh)$.  They also form a BRST quartet as in \qone. In this case
\eqn\qonel{ Q_1 \lth_\ah = - \rhh_\ah,\quad Q_1 \shh^\ah = \oth^\ah,\quad Q_1 \oth^\ah = Q_1 \rhh_\ah = 0 .} 
As above, the minimal BRST transformations of these fields should contain a Lorentz rotation, but this time parametrized by
\eqn\defSh{Ê\Sh_\ah{}^\bh = \l^\g \Oh_{\g\ah}{}^\bh + \lh^\gh \Oh_{\gh\ah}{}^\bh .}
The action of $Q$ on the right-moving non-minimal sector is 
\eqn\QRM{ Q \rhh_\ah = \Sh_\ah{}^\bh \rhh_\bh + \Ah_\ah{}^\bh \lth_\bh,\quad Q \lth_\ah = -\rhh_\ah + \Sh_\ah{}^\bh \lth_\ah ,}   
$$ 
Q \shh^\ah = \oth^\ah + \shh^\bh \Sh_\bh{}^\ah,\quad Q \oth^\ah = - \oth^\bh \Sh_\bh{}^\ah + \shh^\bh \Ah_\bh{}^\ah .$$
The function $\Ah$ is determined by the nilpotency of $Q$ and it turns out to be
\eqn\Ahat{Ê \Ah_\a{}^\b = \l^\g \lh^\dh \Rh_{\g\dh\ah}{}^\bh + \ha \lh^\gh \lh^\dh \Rh_{\gh\dh\ah}{}^\bh .}

The non-minimal contibution to the action is given by
\eqn\nonMinAct{ S_{non-minimal}= Q \int d^2z ( s\Nb\lt + \widehat s \N \widehat\lt),}
which, of course, is BRST trivial. 

\newsec{$AdS_5 \times S^5$ Background}

The action for the type IIB superstring in an $AdS_5\times S^5$
background in the pure spinor formalism can be obtained as follows
\BerkovitsUE. It is assumed that the background supergeometry can be
described in terms of flat currents one-form \MetsaevIT\
\eqn\currents{ J = \ha J^{[\underline{ab}]} M_{[\underline{ab}]} + J^\a
Q_\a
+ J^{\underline{a}} P_{\underline{a}} + J^\ah Q_\ah ,}
where $M, Q, P$ are the generators of the of the $PSU(2,2|4)$ Lie
algebra. Note that $J^{[\underline{ab}]}$ is related to the connection
for the rotations and $J^A = ( J^{\underline{a}}, J^\a, J^\ah )$ are
related to the vielbein of the background geometry. In fact, these
relations are
\eqn\relOE{ J^{[\underline{ab}]} = d Z^M \O_M{}^{[\underline{ab}]},
\quad J^A = d Z^M E_M{}^A ,}
where $\O$ is the spin connection and $E$ is the vielbein of the background.

The flatness condition of the current implies that it can written as
$g^{-1} d g$, where $g$ is a group element of the coset
${PSU(2,2|4)\over{SO(4,1)\times SO(5)}}$. Alternatively, the vielbein
and the connection satisfy certain constraints. In fact, the flatness
condition on $J = dZ^M J_M$ can be written as
\eqn\flat{ \p_{[N} J_{M]} + [ J_N , J_M ] = 0 .}
Using the $PSU(2,2|4)$ Lie algebra, the torsion and the curvature
components are obtained. Recall that the torsion and curvature two-forms are defined as
\eqn\TR{ T^A = d E^A + E^B \O_B{}^A,\quad R_A{}^B = d\O_A{}^B + \O_A{}^C \O_C{}^B ,}
where the wedge product is assumed and the non-vanishing Lorentz
connection elements are given in terms of $\O^{[\underline{ab}]}$ of \relOE\ as
\eqn\nvO{ \O_\a{}^\b = {1\over4} (\g_{[\underline{ab}]})_\a{}^\b
\O^{[\underline{ab}]},\quad \O_\ah{}^\bh = {1\over4}
(\g_{[\underline{ab}]})_\ah{}^\bh \O^{[\underline{ab}]} ,}
where $\g$ are the ten-dimensional symmetric gamma matrices. The
equation \flat\ determines that the non-zero components of the torsion are

\eqn\nvT{ T_{\a\b}{}^{\underline{a}} = - \g^{\underline{a}}_{\a\b},
\quad T_{\ah\bh}{}^{\underline{a}} = - \g^{\underline{a}}_{\ah\bh},
\quad T_{\underline{a}\bh}{}^\a = \ha (\g_{\underline{a}} \eta
)_\bh{}^\a ,\quad T_{\underline{a}\b}{}^\ah = - \ha (\g_{\underline{a}} \eta )_\b{}^\ah ,}
where $\eta_{\a\bh} = (\g^{01234})_{\a\bh}$ and $\eta^{\a\bh} = -
(\g^{01234})^{\a\bh}$. Note that $\eta^2=1$. Similarly, the non-zero
components of the curvature are
\eqn\nvR{ R_{\a\bh}{}^{ab} = - (\g^{ab} \eta )_{\a\bh},
\quad R_{\a\bh}{}^{a'b'} = (\g^{a'b'} \eta )_{\a\bh},\quad
R_{cd}{}^{ab} = \d_c^{[a} \d_d^{b]},\quad R_{c'd'}{}^{a'b'} =
-\d_{c'}^{[a'} \d_{d'}^{b']} ,}
where the index $\underline{a} = ( a , a' )$ with $a$ standing for
the $SO(4,1)$ vector index and $a'$ for the $SO(5)$ vector index.

Note that \nvT\ and \nvR\ satisfy the type II supergravity of
\BerkovitsUE\ constraints involving torsion and curvatures only.
They also imply that the other non-zero background fields are the
Ramond-Ramond field strength $P^{\a\bh}$ and the component $B_{\a\bh}$
of the Kalb-Ramond field. They have the values
\eqn\PB{P^{\a\bh} = - \ha \eta^{\a\bh}  ,\quad B_{\a\bh} = \eta_{\a\bh} .}
Plugging the $AdS_5\times S^5$ background fields in the generic
Type IIB world-sheet action, it becomes \BerkovitsYR\
\eqn\Scomp{ S = \int d^2z ~ \ha J^{\underline{a}} \Jb^{\underline{b}}
\eta_{\underline{ab}} + \ha ( J^\a \Jb^\bh + J^\bh \Jb^\a )
\eta_{\a\bh} + d_\a \Jb^\a + \widehat d_\ah J^\ah  - \ha \eta^{\a\bh} d_\a \widehat d_\bh }
$$ + \o_\a \Nb \l^\a + \oh_\ah \N \lh^\ah + \ha ( N^{ab} \widehat N_{ab} - N^{a'b'} \widehat N_{a'b'} ) ,$$
where
$$ \Nb \l^\a = \p \l^\a + {1\over4} \l^\b \pb Z^M \O_{M\b}{}^\a,\quad   \N \lh^\ah = \p \lh^\ah  + {1\over4} \lh^\bh \p Z^M \O_{M\bh}{}^\ah ,$$
$$ N^{\underline{ab}} = \ha (\l \g^{\underline{ab}} \o),\quad \overline N^{\underline{ab}} = \ha (\lh \g^{\underline{ab}} \oh) .$$

In an useful notation, introduced in \BerkovitsXU,  the world-sheet fields takes values in the $PSU(2,2|4)$ super Lie algebra. They are
\eqn\wsf{ J_0 = \ha J^{\underline{ab}} M_{\underline{ab}},\quad J_1 = J^\a Q_\a,\quad J_2 = J^{\underline{a}} P_{\underline{a}},\quad J_3 = J^\ah Q_{\ah} ,}
$$ \l = \l^\a Q_\a,\quad \o = \o_\a \eta^{\a\bh} Q_\bh,\quad \lh = \lh^\ah Q_\ah,\quad \oh = \oh_\ah \eta^{\b\ah} Q_\b,\quad d = d_\a \eta^{\a\bh} Q_\bh,\quad \widehat d = \dhh_\ah \eta^{\b\ah} Q_\b .$$
In this way, the covariant derivatives for the pure spinor variables $\l$ and $\lh$ are expressed as
\eqn\covder{ \Nb \l = \pb \l + [ \Jb_0 , \l ] = \Nb \l^\a Q_\a,\quad \N \lh = \p \lh + [ J_0 , \lh ] = \N \lh^\ah Q_\ah .}
Similarly, the covariant derivative of any field has the form $\N\cdot = \p\cdot + [ J_0 , \cdot ]$.

The normalization for the supertraces of the $PSU(2,2|4)$ is chosen to be
\eqn\normstr{ \Str ( P_{\underline{a}} P_{\underline{b}} )= \eta_{\underline{ab}},\quad \Str ( Q_\a Q_\bh ) = -2 \eta_{\a\bh} ,}
$$ \quad \Str ( M_{ab} M_{cd} ) = - \eta_{a[c} \eta_{d]b},\quad \Str ( M_{a'b'} M_{c'd'} ) =  \eta_{a'[c'} \eta_{d']b'} ,$$
and the other possible products of two generators have vanishing supertrace. With the notation \wsf\ and the result \normstr, the action \Scomp\ can be written as
\eqn\action{ S = \left< \ha J_2 \Jb_2 + {1\over4} J_1 \Jb_3 -
{1\over4} J_3 \Jb_1 - \ha d \Jb_1 + \ha \dhh J_3 +{1\over4} d \dhh + \ha \o \N \l - \ha
\oh \N \lh -  {1\over4} N \widehat N \right> ,}
where $\left< \cdots \right> = \int d^2 z \Str ( \cdots )$ and
\eqn\Ns{ N =  \{ \l , \o \} ,\quad \widehat N = - \{ \lh , \oh \} .}

\subsec{BRST transformations}
The quantization in the pure spinor formalism is performed through a postulated BRST charge
\eqn\Qbrst{ Q = \oint \l^\a d_\a + \lh^\ah \widehat d_\ah = - \ha \oint \Str ( \l d + \lh \widehat d ) .}
The BRST transformation of the world-sheet fields will be obtained now. The idea is to define a canonical momentum conjugate of the variable $Z^M$ and impose canonical Poisson brackets between the pairs of canonical variables. Namely, the pairs $(\l , \o)$ and $(\lh , \oh)$ are canonical conjugate pairs.  The momentum conjugate of $Z^M$ is defined as
\eqn\momZ{ P_M = {\d S\over{\d \p_0 Z^M}} ,}
where $\p = \p_0 - \p_1$ and $\pb = \p_0 + \p_1$. For the action \Scomp, the momentum becomes
\eqn\PM{ P_M = - E_M{}^\a d_\a - E_M{}^\ah d_\ah + \p_0 Z^N E_N{}^{\underline{a}} E_M{}^{\underline{b}} \eta_{\underline{ab}} + (-1)^{M+1} \p Z^N E_N{}^\bh E_m{}^\a \eta_{\a\bh} }
$$ + (-1)^{M+1} \p_1 Z^N E_N{}^\a E_M{}^\bh \eta_{\a\bh} + \O_{M\a}{}^\b \l^\b \o_\b +  \O_{M\ah}{}^\bh \lh^\bh \oh_\bh .$$
Note that, it is possible to express $d$ and $\widehat d$ in terms of canonical variables, then it is possible to compute the BSRT transformation of any field $\Psi$ as
\eqn\Qpsi{ Q \Psi = \oint [  \l^\a d_\a + \lh^\ah \widehat d_\ah , \Psi ] ,}
where the equal world-sheet times canonical commutation relations
\eqn\commc{ [ P_M(\s) , Z^N(\s') ] = - \d_M^N \d(\s-\s'),}
$$ [ \l^\a(\s) , \o_\b(\s') ] = \d_\b^\a \d(\s-\s'),\quad [ \lh^\ah(\s) , \oh_\bh(\s') ] = \d_\bh^\ah \d(\s-\s') ,$$
have to be used.

Consider first the group element $g$. The only non-zero commutator in \Qpsi\ comes from the commutator between $P_M$ and $g$. The result is
\eqn\Qg{ Q g = g ( \l + \lh + \S ) ,}
where
\eqn\defS{ \S = \ha \S^{\underline{ab}} M_{\underline{ab}} = \ha ( \l^\a \O_\a{}^{\underline{ab}} + \lh^\ah \O_\ah{}^{\underline{ab}} ) M_{\underline{ab}} ,}
is a rotation of $g$ with a field-dependent parameter and it was ignored in \BerkovitsXU\ but it is relevant to achieve nilpotency of $Q$. Note that the same property appeared in the heterotic string in a generic background \ChandiaIX. The BRST transformation of the other world-sheet fields are easy to compute, except for $d$ and $\dhh$. The complication comes from the commutators between the $d$ and $\widehat d$ in \Qpsi. The result is
\eqn\Qothers{ Q \l = - \{ \l , \S \},\quad Q \o = d - \{ \o , \S \},\quad Q \lh = - \{ \lh , \S \},\quad Q\oh = \widehat d - \{ \oh , \S \} .}
$$
Q d = - 2 [ J_2 , \l ] + [ N ,\lh ] + [ d , \S ] ,\quad Q \widehat d = 2 [ \Jb_2 , \l ] - [ \Nh , \l ] + [ \widehat d, \S ] .$$
Note that $Q J_3$ is proportional to $Q d$ on-shell. Similarly, $Q \Jb_3$ is proportional to $Q \widehat d$.

Now it will be verified that $Q^2 g = 0$. Acting with $Q$ on \Qg\ and using \Qothers\ it is obtained that
\eqn\QQg{  Q^2 g = g ( Q \S + \S^2 + \{ \l ,\lh \} ) ,}
where $\l^2=\lh^2=0$ because they are pure spinors. It remains to prove that
\eqn\QS{  Q \S + \S^2 + \{ \l ,\lh \} = 0 ,}
to verify nilpotency of $Q$ on $g$. Note that the lhs of \QS\ can be written as
\eqn\QSp{ \ha ( Q \S^{\underline{ab}} - \S^{\underline{a}}{}_{\underline{c}} \S^{\underline{cb}} ) M_{\underline{ab}} + \{ \l , \lh \} .}
When expressed in terms of components, \QSp\ contains terms with $\l^\a \l^\b$, $\lh^\ah \l^\bh$ and $\l^\a \lh^\bh$. The anticommutator between the pure spinor variable will contribute to the terms with $\l^\a \lh^\bh$  only. Consider the term with $\l^\a \l^\b$ in \QSp. It is
$$
\l^\a \l^\b  \left( \p_\b \O_\a{}^{\underline{ab}} - \O_{\b\a}{}^\g \O_\g{}^{{\underline{ab}}} - \O_\b{}^{\underline{a}}{}_{\underline{c}} \O_\a{}^{{\underline{cb}}} \right) M_{\underline{ab}} .$$
Symmetrizing in $_{\a\b}$, this combination becomes proportional to $R_{\a\b}{}^{\underline{ab}}$ which vanishes in $AdS_5\times S^5$ background. Considering the term with $\lh^\ah \l^\bh$ produces the curvature $R_{\ah\bh}{}^{\underline{ab}}$ which also vanishes. Consider the terms with $\l^\a \lh^\bh$in \QSp. It is proportional to
$$\l^\a \lh^\bh \left( R_{\a\bh}{}^{\underline{ab}} M_{\underline{ab}} + (\g^{ab} \eta)_{\a\bh} M_{ab} - (\g^{a'b'}\eta)_{\a\bh}  M_{\underline{a'b'}} \right) .$$
This expression vanishes when we plug the values of the curvature
\nvR. Therefore, the equation \QS\ is satisfied and consequently, $Q^2
g$ is zero. It is interesting to note that the equation \QS\ can be
written in the form
\eqn\flatA{ (Q +{\cal A})^2=0,}
where ${\cal A} = \S + \l + \lh$, so we interpret ${\cal A}$ as the
flat connection for the BRST differential.

In summary, the action for the minimal variables is given by \action\ and the BRST transformations for the world-sheet variables are given by \Qg\ and \Qothers.

We are interested in including the non-minimal fields
$(\lt,\ot,r,s,\lth,\oth,\rhh,\shh)$
in the action \action.  These non-minimal fields cannot change
the cohomology of $Q$, so their modification has to be in the so-called BRST
quartet
\eqn\qone{Q_1 =-\half \oint d\sigma\,\Str( \ot r + \oth \rhh)}
This defines how $Q_1$ acts on the non-minimal fields
\eqn\qoneaction{ Q_1 \lt = -r,\quad Q_1 s = \ot,\quad Q_1 \ot =0,\quad
Q_1 r = 0,}
$$Q_1 \lth = -\rhh,\quad Q_1 \shh = \oth,\quad Q_1 \oth =0,\quad
Q_1 \rhh = 0.$$

We will now  call the original BRST \Qbrst\ $Q_0$ and the new BRST
charge is
\eqn\newbrst{Q = Q_0 + Q_1.}
Consistency of $Q$ implies that
\eqn\qzeroqone{Q_0^2+ \{Q_0,Q_1\} +Q_1^2=0}
has to be true for all fields. $Q_1$ only acts on the non minimal fields
and will not be modified. We will use this equation to find how
$Q_0$ acts on the non-minimal fields \ChandiaIMA. However it will not be possible
to demand $Q_0^2=0$ on the nonminimal fields. Let us start with
$r$. Since $Q_1 r=0,$, we have to solve
\eqn\brstrA{(Q_0  + Q_1)Q_0 r = 0.}
We will suppose a solution of the form
\eqn\brstrB{Q_0r = [ A,r] +\{B,\ot\} +
[C,\rhh]+\{D,\oth\} + [E,\lt]+ [F,\lth],  }
where $(A,B,C,D,E)$ are functions of only the minimal fields. We suppose
the commutator form since the right hand side must be algebra
valued. We can now follow the same procedure as in Section 3. Since $r$ has 
vanishing conformal dimension and there are no
negative dimension fields $B$ and $D$ must
vanish. As a simplifying assumption, we will set $C=F=0$ and check
that this is consistent. Since all BRST transformations have a field
dependent Lorentz transformation we will set $A= -\S$. To find $E$ we use the same 
strategy as before. First we note that
$Q_1$ constraints the ghost numbers of $r$ and $\lt$ to satisfy $q_r - q_\lt=1$.
This implies that $q_E=2$. The fact that $r$ and $\lt$ have the same
${Z}_4$ charge means $E$ has to have vanishing $Z_4$ charge. These two
requirements hint that $E$ is proportional to $\{\l,\lh\}$. We also
have to guess $Q_0\lt$. Using the same reasoning we find there is
no contribution to it besides $\{\S,\lt\}$.  Inserting 
these guesses in \brstrA\ and using \flatA\ we find that $E$ is simply $\{\l,\lh\}$.

This strategy can be used to all non-minimal fields. The final result is
\eqn\nonminimalbrst{
Q\lt=-r -\{\S,\lt\},\quad Qs=\ot -\{\S,s\},}
$$Q\ot=-\{\S,\ot\} - [\{\l,\lh\},s],\quad Qr = -\{\S,r\}+[\{\l,\lh\},\lt],$$
$$Q\lth=-\rhh -\{\S,\lth\},\quad Q\shh=\oth -\{\S,\shh\},$$
$$Q\oth=-\{\S,\oth\}- [\{\l,\lh\},\shh],\quad Q\rhh = -\{\S,\rhh\}+
[\{\l,\lh\},\rhh].$$

Again, the action for the non-minimal variables is given by the BRST exact form 
\eqn\actionNM{ÊS_{non-minimal} = -\ha Q \left< s \Nb \lt + \shh \N \lth \right> .}
Using the minimal and non-minmal BRST transformations, this action becomes
\eqn\snmev{ ÊS_{non-minimal} = -\ha \left< \ot \Nb \lt - s \Nb r + \oth \N \lth - \shh \N \rhh \right> }
$$ -\ha \left< \Jb_1 \{ \lh , [ \lt , s ] \} + \Jb_3 \{ \l , [ \lt , s ] \} + J_1 \{ \lh , [ \lth , \shh ] \} + J_3 \{ \l , [ \lth , \shh ] \} \right>  .$$

\newsec{Conclusion}

In this work we studied the the non-minimal formulation of the pure spinor string in general curved backgrounds. We derived 
the coupling of the non-minimal fields to the background and minimal variables demanding consistency of the BRST charge. 

The next step is to construct the composite $b$ ghost in a general curved background. In order to do this, two possible 
path could be taken. The method used in  \ChandiaIMA\ is the curved space version of the BRST hierarchy of tensor fields 
originally used in \refs{\BerkovitsPX,\BerkovitsVI}. However, as it can be seen in \ChandiaIMA, this method becomes very cumbersome 
in a curved background. For a general Type II background, or even the simpler case of $AdS_5\times S^5$, the tensor hierarchy 
equations become even more difficult to solve due to the coupling of left- and right-moving ghosts in the curvature term 
of the BRST trasformation. The more promissing way is to used the new fermionic composite fields $(\Gamma^m,\overline\Gamma^m)$ introduced in \BerkovitsPLA. Using $\overline\Gamma^m$, a simplified expression for $b$ can be obtained. This was the approach used in 
\BerkovitsAMA\ where the complete $b$ ghost was found. We hope to generalize this construction for Type II strings. 

Understing better the non-minimal fields in curved backgrounds can also be useful for tree level quantum computations. One 
of the uses of the non-minimal fields is to regularize the zero mode integration in quantum expectation values \BerkovitsBT. 
The integration measure found in \BerkovitsBT\ can be reduced to the original pure spinor measure  \BerkovitsFE\ in a flat background, but this depends on the existance of the zero modes. In a curved background, the coupling to the spin connection may generate 
a mass term, which means not all components might have zero modes. For this reason, quantum expectation values in curved spaces 
might be better defined using the prescription in \BerkovitsBT. Besisdes amplitudes, another observable that in principle can be 
computed using this measure is the string duals to Wilson loops in $AdS_5\times S^5$. We plan to look into this problem in the 
future. 

As a side note, it could be interesting if the BRST flat connection ${\cal A} = \S + \l + \lh$ in eq. \flatA\ could be used as an alternative way to define physical fluctuations which are not scalars in $AdS$. For instance, a physical fluctuation will generate a change in ${\cal A}$, and the linearized equation of motion for this variation is $ \{ Q+ {\cal A} , \delta{\cal A} \}=0$. The ghost number one cohomology of $Q+{\cal A}$ could be related to physical vertex operators. Note that the ghost number one cohomology of $Q$ is given by the conserved currents related to the global symmetries of the $AdS$ action \MikhailovRX. 

\bigskip

\noindent
{\bf Acknowledgements:} 
 We would like to thank William D. Linch III for useful discussions. The work of OC and BCV is partially supported by 
 a FONDECYT grant number 1120263.

 \listrefs
 
\end